\documentstyle[epsfig, twocolumn, 12pt]{esapub}

\begin{document}

\setlength{\parindent}{0pt}
\setlength{\parskip}{ 10pt plus 1pt minus 1pt}
\setlength{\hoffset}{-1.5truecm}
\setlength{\textwidth}{ 17.1truecm }
\setlength{\columnsep}{1truecm }
\setlength{\columnseprule}{0pt}
\setlength{\headheight}{12pt}
\setlength{\headsep}{20pt}
\pagestyle{esapubheadings}

\title{\bf SOURCE COUNTS AND BACKGROUND RADIATION}

\author{ {\bf A.~Franceschini$^1$, H. Aussel$^2$, A.~Bressan$^3$, 
C.J. Cesarsky$^2$, L.~Danese$^4$, } \\
{\bf G.~De Zotti$^3$, D. Elbaz$^2$, G.L.~Granato$^3$, P.~Mazzei$^3$, 
L.~Silva$^4$}
\vspace{2mm} \\
$^1$Dipartimento di Astronomia di Padova, I-35122 Padova, Italy\\
$^2$Service d'Astrophysique, CEN, Saclay, France \\
$^3$Osservatorio Astronomico di Padova, I-35122 Padova, Italy\\
$^4$S.I.S.S.A., Trieste, Italy \\
}

\maketitle

\begin{abstract}

Our present understanding of the extragalactic source counts and 
background radiation at infrared and sub-mm wavelengths is reviewed.
Available count data -- coming in particular
from a very deep survey by ISO in the near- and mid-IR and from deep
IRAS surveys at longer wavelengths -- are used to constrain evolutionary
models of galaxies and Active Nuclei. \\
The extragalactic IR background radiation (CIRB), on the other hand, provides
crucial information on the integrated past IR emissivity, including
sources so faint to be never accessible. 
Two spectral bands, the near-IR and sub-millimeter -- where local foregrounds 
are at the minimum -- are suited to search for the CIRB. These cosmological 
windows are ideal to
detect redshifted photons from the two most prominent broad emission
features in galaxy spectra: the stellar photospheric 1 micron peak and
the one at 100 micron due to dust re-radiation. 
The recently claimed detection of an isotropic diffuse component at 
$\lambda\simeq 100-200\ \mu m$ would support, whenever confirmed, 
the evidence for strong cosmic
evolution from $z=0$ to $z\sim 2$ of faint gas-rich late-type galaxies, 
as inferred from direct long-$\lambda$ counts. 
Furthermore, an equally intense CIRB flux measured in the wavelength range
200 to 500 $\mu m$ may be in support of models envisaging a 
dust-enshrouded phase during formation of spheroidal galaxies 
(e.g. Franceschini et al., 1994):
the background spectrum at such long wavelengths would indicate 
an high redshift ($z>2-3$) for this active phase of star formation, implying 
a relevant constraint on structure formation scenarios. 
\\
We argue that, given the difficulties for sky surveys in this spectral domain,
only sporadic -- and probably inconclusive -- tests of these ideas will be 
achieved in the next several years. Only a mission like FIRST, in combination 
with large mm arrays and optical telescopes, will allow to 
thoroughly investigate, via photometric and spectroscopic surveys, these
IR-bright early phases in the evolution of galaxies and AGNs.
\vspace {5pt} \\


  Key~words: source counts; evolution; background radiation.

\end{abstract}

\section{INTRODUCTION}

Remarkable progresses have been recently made in the exploration of the
high-redshift universe at optical and near-IR wavelengths. 
A combined effort of
high spatial resolution (e.g. with the HST), large collecting power (e.g. 
with the Keck telescope)
and suitable selection techniques (e.g. the Lyman "drop-out"), have allowed
to discover and characterize galaxian units up to redshifts of the order
of 5. 

Beyond the K-band spectral limit and up to 20 $\mu m$, the ISO mid-IR camera
has started to complement the optical data with a sensitive coverage in
this previously mostly unexplored domain.

Finally, the three decades in frequency longward of 20 $\mu m$ up to the
radio band will be somehow tackled by various missions and observational 
campaigns, but are likely to remain a poorly explored field in the foreseable 
future. In particular ISO, which is observing here with the
long-wavelength camera PHOT, has a limited sensitivity due to the small
primary collector and the correspondingly high confusion noise, but 
similar problems are also likely to limit SIRTF at $\lambda > 50 \ \mu m$.
Bolometer arrays on large millimetric telescopes are powerful 
instruments, but confined to quite small sky coverages because of 
the small number of detector elements and very long integration times 
needed to overcome the atmospheric 
noise even in the 2-3 cleanest spectral windows.

In spite of these and other efforts, exploring the distant universe will 
keep increasingly difficult at increasing $\lambda$ for many years to 
come. It is then worthwhile to review here the possible impact a mission 
like FIRST would have in this context.

Apart from useful constraints on specific galaxy evolutionary scenarios 
devised to coherently interprete an exceedingly large
dataset, a general lesson we can infer from available data on distant
galaxies is that {\sl results in a given waveband may be very hard to
reproduce at other wavelengths} and that {\sl properties of galaxy
emissivity and evolution are strongly dependent on the selection
wavelength}. 

So, the assessment of FIRST's capabilities in the cosmological context 
has inevitably to deal with somewhat model-dependent and indirect 
arguments. Nevertheless, and to anticipate our main conclusions, we 
argue that ESA's cornerstone mission FIRST will have a profound impact 
on cosmology and will address problems which can be tackled in no other 
way.

The paper is organized as follows. 
The present status of the observations and interpretations of faint galaxy 
samples in the IR are summarized in Sects. 2 and 3, while the complementary
constraints set by observations of the IR background are mentioned in Sect. 4.
A tentative scheme of galaxy evolution, discussed in Sect. 5, is used
in Sect. 6 to predict and discuss the impact of FIRST in the 
exploration of the distant universe, and to compare it with other future 
far-infrared and sub-millimetric facilities.

\section{GALAXY COUNTS IN THE IR: CURRENT OBSERVATIONAL STATUS}

The number-flux relationship of cosmological sources -- combined with 
information on the local 
luminosity functions and the spectral K- and evolutionary corrections --
is a classical test of evolution of the source emissivity
with cosmic time.

 \begin{figure}[!ht]
  \begin{center}
    \leavevmode
    \centerline{\epsfig{file=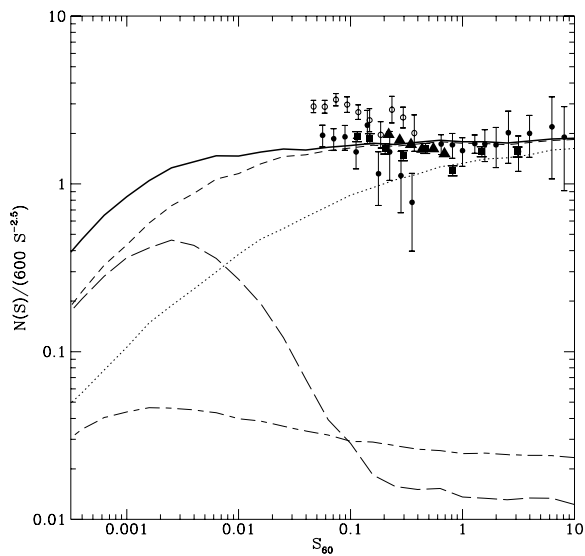,width=8.0cm}}
\vspace{-1.cm}
  \end{center}
  \caption{\em Differential galaxy counts of galaxies at 60 $\mu m$
normalized to the euclidean law $600 S^{-2.5}\ sr^{-1}/Jy{-1}$.
The sources of the data are mentioned in Sect. 2.1. 
The dotted line correspond to the predicted no-evolution case (Sect. 
3.1). The short-dashed line is
the predicted counts for gas-rich evolving galaxies (including
spirals, starbursts, irregulars. Long-dashes correspond to dust-emission
from early-types (E/S0) during an early evolution phase. Long-short dash
are AGNs. More details on the models are provided in Sect. 3.3.
}
  \label{fig1}
\end{figure}

With the aim of constraining galaxy evolution schemes to be used in the
following Sects., we briefly summarize here the presently available scanty 
information on source counts in the infrared.

\subsection{Counts based on the IRAS survey}

The all-sky survey by IRAS has revealed a numerous local population of 
luminous dust-rich star-forming galaxies at $z<0.1$. But, in addition, 
in a repeatedly scanned fraction of the sky IRAS has reached flux limits 
faint enough to start detecting galaxies at cosmological distances.

Faint number counts of galaxies at 60 $\mu m$ (where Galactic cirrus
contamination is at the minimum and survey sensitivity at the maximum) 
have been discussed by Hacking \& Houck (1987), Gregorich et al. (1995),
Bertin, Dennefeld, Moshir (1997), while the bright-flux normalization
to local source populations was based on analyses of the PSC by 
Rowan-Robinson et al. (1991). A collection of 60 $\mu m$ differential counts
is reported in Figure 1. As shown there, the counts at faint 
fluxes appear in excess of no-evolution extrapolations from the 
bright flux limits.

Extensive efforts of optical identifications and spectroscopic follow-up
by Saunders et al. (1990, 1997), Lonsdale et al. (1990),
Oliver et al. (1995) support conclusions in favour of evolution 
(even a strong one) by  Franceschini et al. (1988), 
Lonsdale et al., Oliver et al.

However, as the few most distant galaxies in the faintest samples are found 
at $z\simeq 0.2-0.3$, any conclusions based on IRAS are to be taken as 
tentative only, large-scale inhomogeneities possibly affecting such 
shallow samples.

\subsection{Near- and mid-IR counts by ISO-CAM}

The Infrared Space Observatory has recently allowed some very significant, 
though preliminary, steps forward. The occasion was a long integration 
with the well-performing mid-IR camera ISOCAM in two filters 
(LW2=5-8.5 $\mu m$, 
LW3=12-18 $\mu m$) on the very extensively studied Hubble Deep Field
(Rowan-Robinson et al., 1997).
Redundancy in the number of elementary integrations per sky pixel have 
allowed to reach the faintest limiting fluxes ever at these $\lambda$,
three decades fainter than IRAS at 12 $\mu m$.

 \begin{figure}[!ht]
  \begin{center}
    \leavevmode
    \centerline{\epsfig{file=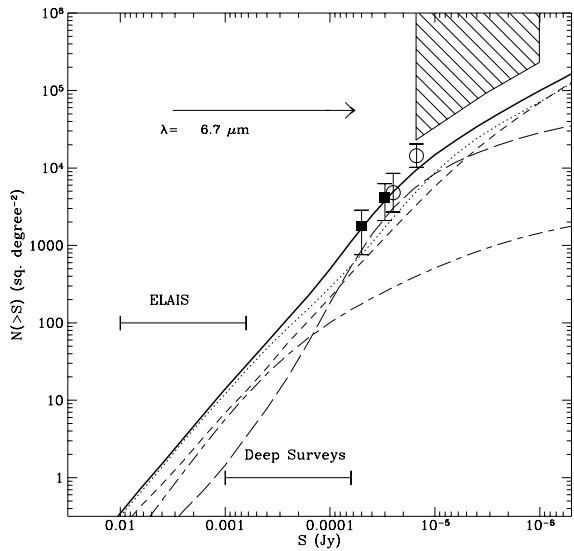,width=8.0cm}}
  \vspace{-1.cm}
  \end{center}
  \caption{\em Integral galaxy counts of galaxies at 6.7 $\mu m$ based 
  on the CAM-LW2 survey in the HDF (see Oliver et al. for more details).
  Meaning of the lines as in Fig. 1. The horizontal line marks the areal
  density corresponding to a confusion-limited survey. See Oliver et al. 
for more details.
}
  \label{fig2}
\end{figure}

Source counts at the effective wavelengths $\lambda$=6.7 and 15 $\mu m$ are 
discussed by Oliver et al. (1997) and results appear in Figures 2 and 3.
Number of detections and sky coverage are 27 sources/5 sq.arcmin and 
22 sources/15 sq.arcmin, respectively.
As can be seen, the source confusion limit for the ISO 60-cm primary collector
is almost reached at 15 $\mu m$.
A refined analysis of these data, based on a multi-scale wavelet transform
(Stark et al., 1997), is reported by Aussel et al. (1997).

Various other similarly deep CAM surveys both in GT and OT are being performed
over more significant sky areas (see C. Cesarsky, these Proceedings), 
while a shallow survey over 20 square degrees is being performed by the
European Large Area ISO Survey (ELAIS, M. Rowan-Robinson, these Proceedings) 
over 20 square degrees.

 \begin{figure}[!ht]
  \begin{center}
    \leavevmode
    \centerline{\epsfig{file=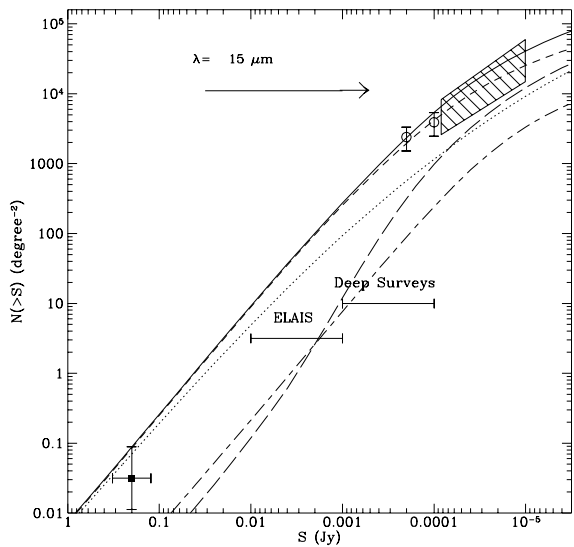,width=8.0cm}}
\vspace{-1.cm}
  \end{center}
  \caption{\em Integral galaxy counts at 15 $\mu m$ based on the CAM-LW3
survey in the HDF.
  The shaded region corresponds to a boundary on the counts coming from an 
  analysis of the background fluctuations in the ISO LW3 map. 
}
  \label{fig3}
\end{figure}

A potentially useful match with these data comes from the extensive 
ground-based surveys in the K band. Figure 4 summarizes some 
results on galaxy counts.

\subsection{Far-IR counts by ISO-PHOT }

While the ISO long-wavelength camera C100 is still under test, the C200
camera covering the 120-240 $\mu m$ range is already successfully and
rutinely performing.
A deep survey has been performed by Kawara et al. (these Proceedings) 
with PHT C200 at $\lambda=175\ \mu m$ over
22'x22' in the Lockman Hole (at the absolute minimum of the Galactic 100
$\mu m$ emission).

Probable extragalactic sources, unrelated to "cirrus" emission, 
have been found, with preliminarly calibrated fluxes close to or below
$100\ mJy$ (see in Figure 5 the corresponding counts). 
Optical counterparts are either faint or absent on POSS, 
indicative of remote galaxies. Indeed we expect that the positive 
K-correction at such long-$\lambda$ operates to favour selection of 
fairly high-z with respect to low-z galaxies.

Similar ISO-PHOT surveys over larger sky areas are being performed by 
Kawara et al. and Puget et al.

 \begin{figure}[!ht]
  \begin{center}
    \leavevmode
    \centerline{\epsfig{file=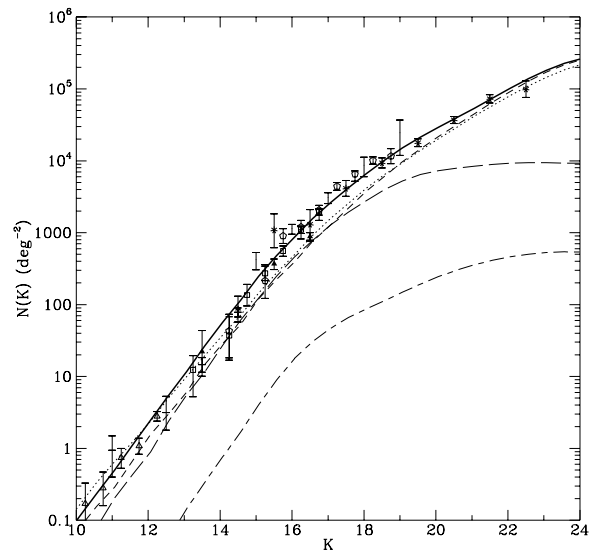,width=8.0cm}}
\vspace{-1.cm}
  \end{center}
  \caption{\em Differential galaxy counts in the K band ($\lambda= 2.2 \mu m$). 
Data collected by Gardner et al. (1995).
Meaning of the lines as in the previous figures. 
}
  \label{fig4}
\end{figure}

\section{INTERPRETATIONS OF THE IR COUNTS}

As previously mentioned, thanks to ISO we are presently in
the course of a real quantum jump in our knowledge of distant IR
sources. Only a tiny fraction of ISO survey data are presently analyzed,
much more information will be available in one year from now.
Nevertheless, it is of interest to try to infer from the present data
constraints on galaxy evolution to be compared with the optical data and
suitable to discuss the possible impact of future missions. 

\subsection{The $\lambda$-dependent local luminosity function}

The local luminosity function (LLF) is a fundamental constraint on 
the evolutionary history of galaxy emissivity.
In the IR-mm, its knowledge mostly relies on the IRAS all-sky surveys
and on wide-area surveys in the K-band.
We used galaxy LLF by Saunders et al. (1990) at 60 $\mu m$, by Rush 
et al. (1993) at 12 $\mu m$ and a recent evaluation by Gardner et al. 
(1997) at 2.2 $\mu m$.  Within the envelopes provided by these global LLFs,
further subdivision into different morphological classes is made following
Franceschini et al. (1988b).

The $\lambda$-dependent LLF has been extrapolated to $\lambda > 100\ \mu m$
by Franceschini, Andreani \& Danese (1997) using
$1.3\ mm$ observations of a complete sample of 30 bright IRAS galaxies
selected at 60 $\mu m$, and convolving the IRAS LLF with a IR-mm 
bivariate luminosity distribution.
This is very critical for the interpretation of source counts
and background data at long wavelengths.
We note, however, that this procedure could potentially miss 
some faint dust-rich and cold galaxies. $Mm$-selected galaxy samples, 
impossible to get before the PLANCK Surveyor and the FIRST
mission, would be needed to prove this.

 \begin{figure}[!ht]
  \begin{center}
    \leavevmode
    \centerline{\epsfig{file=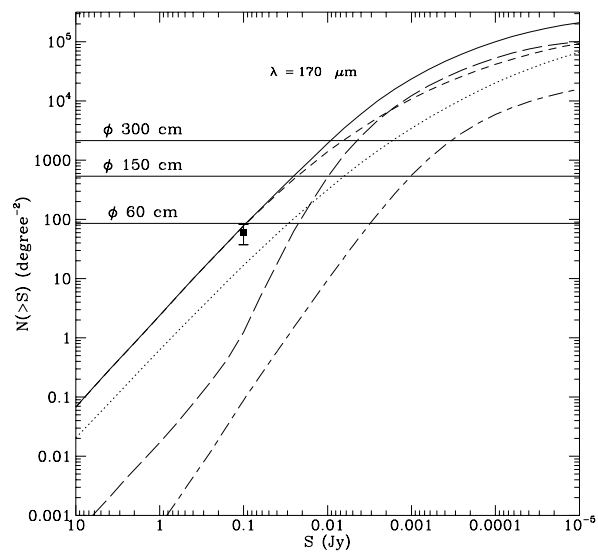,width=8.0cm}}
\vspace{-1.cm}
  \end{center}
  \caption{\em Galaxy integral counts at $\lambda= 170\ \mu m$. The
data-point is based on Kawara et al. (see Sect. 2.3). 
Meaning of various lines as in previous figs.   
The orizontal lines mark the onset of confusion as a function of the primary
collector's diameter.   Note the importance of the latter parameter for
space missions at these wavelengths. According to this, a 3.5m FIRST is 
not expected to be confusion limited above a flux $S_{170}\sim 5 mJy$.
}
\label{fig5}
\end{figure}

Substantial contributions by ISO are soon expected. In particular at 
$\lambda=170\ \mu m$ the ISOPHOT Far-IR Serendipity Survey 
(Bogun et al. 1996), is covering in the ISO slew mode about 
15\% of the whole sky and detecting some $\sim 3000$ galaxies to $S_{170}>\ 
1\ Jy$. ELAIS (see above) is surveying at $\lambda=95$ and 15 $\mu m$ 
twenty square degrees to $\sim 20\ mJy$ and 1 mJy, respectively, 
while additional smaller areas are being mapped by the Guaranteed Time
Shallow Survey (C. Cesarsky, these Proceedings).
At least hundreds of local galaxies are expected at each $\lambda$.

Knowledge of the local IR volume emissivity of galaxies
allows a zero-th order evaluation of possible evolutionary effects 
traced by the observed number counts.
The differential counts per sr at a given flux $S$ are given by:
\begin{equation}
{dN \over dS} = 
\int_{z_l}^{z_h}\,dz\,{dV\over dz}\,{d \log L(S,z)\over dS}\, 
\rho[L(S,z),z] \label{eq:dNdS}
\end{equation}
where $\rho[L(S,z),z]$ is the epoch-dependent luminosity function and 
$dV/dz$ is the differential volume element (per unit solid angle).
%
Flux $S$ and rest-frame luminosity $L$ are related by 
\begin{equation}
S_{\Delta \nu} = {L_{\Delta\nu} K(L,z) \over 4\pi d_L^2}, \label{eq:S}
\end{equation}
where $d_L$ is the luminosity distance and
$K(L,z)=$ \\ $(1+z) {L[\nu (1+z)]\over L(\nu)} $ the K-correction.  
A no-evolution evaluation is made assuming $\rho[L(S,z),z]$ = 
$\rho[L(0),z=0]$ and $z_h=1$.

As shown in Figs. 1 to 5 (dotted lines), comparison with the available
counts of predictions based on no-evolution models already 
shows a remarkable dichotomy. While very moderate 
evolution is indicated in LW2 (5-8.5$\mu m$) and in K-band, much
stronger evolution effects with redshift are required to explain the
observed ISOCAM counts in LW3 (12-18 $\mu m$), the IRAS 60 $\mu m$ and
the ISO-PHOT 170 $\mu m$ counts.

\subsection{IR galaxy spectra and optical morphology}

Essential information to interprete the observed number counts may be 
inferred from optical identifications of ISO-HDF and of IRAS deep samples.

A useful guideline for our discussion is to split the IR domain in two parts, 
according to the expected dominant astrophysical contributor:
\\
{\it a)} the near-IR domain (1-8.5 $\mu m$), dominated by starlight emission, 
and including K-band and ISO-CAM LW2;
\\
{\it b)} the far-IR part (10-1000 $\mu m$), dominated by dust reprocessing
(by warm dust, PAH, small grains, and cooler dust at $\lambda \sim
100\ \mu m$), and including ISO-CAM LW3, ISO-PHOT and IRAS 60 $\mu m$.
	    
Galaxy samples selected in the near-IR show composite morphologies, including
early-types (E/S0), dominated by photospheric emission from old 
stellar populations, and late-type, gas-rich systems (Sa through Irregulars).
Sources in the ISO-HDF sample, in particular, have been identified 
by Mann et al. (1997).
The optical through IR broad-band spectrum of a $z=1$ elliptical is shown in 
Figure 6. The spectrum is fitted by an old (t=4 Gyr)
and already massive ($M=4\ 10^{11}\ M_\odot$) stellar population at that 
redshift. As for the morphology, in the WFC high-resolution image the galaxy 
looks as a plain, very regular system, in no way dissimilar to local massive
ellipticals.
This sets an interesting lower limit to the age of the universe
and constrains the formation of very massive field galaxies.

 \begin{figure}[!ht]
  \begin{center}
    \leavevmode
    \centerline{\epsfig{file=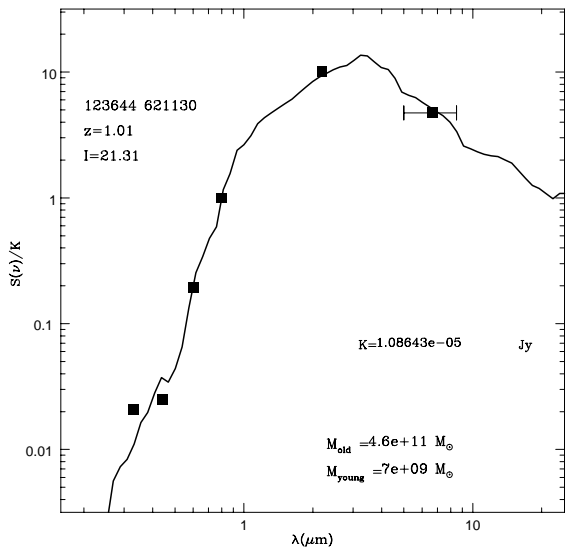,width=8.0cm}}
\vspace{-1.cm}
  \end{center}
  \caption{\em Broad-band optical-IR spectrum of a z=1 elliptical galaxy in
the ISO-HDF sample. See Mann et al. (1997) for the optical and LW2 fluxes. 
The K-band flux comes from the KPNO IRIM image. The model assumes a 4 Gyr old
stellar population with a mass of $4\ 10^{11}\ M_\odot$ plus the
contribution of a younger (1 Gyr) component including 2\% of the barionic 
mass.
}
\label{fig6}
\end{figure}

More general conclusions are drawn with reference to the source list by 
Aussel et al. (1997). Sources identified with E/S0's are typically found at 
$z\sim 1$, because of the combined effect of K-correction plus some moderate
[passive] evolution.
Objects identified as Sp/Ir's are found at lower redshifts ($z=0$ to 0.5). 
    
On the contrary, the small far-IR sample selected by CAM-LW3 and identified
in the HDF includes exclusively 
gas-rich systems (Spirals/Irregulars and Starbursts) at substantial
redshifts (z=0 to $\sim 1$).
Studies of faint 60 $\mu m$ galaxies and hyper-luminus IRAS
galaxies also emphasize normal spirals at moderate
luminosities/redshifts and of starbursts at higher L,z values.

\subsection{IR galaxy counts: interpretation}

Moderate evolution is indicated by galaxy counts in the 
near-IR domain ($1<\lambda<8.5\ \mu m$). This is fully consistent with
purely passive evolution at redshifts less than a few to several for the E/S0 
population: the brighting is of typically less than 1 mag by z=1 to 1.5.
Similar conclusions are derived for galaxies in high-z clusters, but it
would be essential to test them in the field population to understand if there
are any significant differences.

The other component, the gas-rich systems, are also bound 
to evolve very slowly at these wavelengths, again 
consistent with the evidence that the flux is mostly contributed by old 
stellar populations whose mass and luminosity slowly evolve with cosmic 
time. These recipes (i.e. pure luminosity evolution, $z_{form}$=2.5 
and $z_{form}$=7 for late- and early-type galaxies, as discussed in 
Franceschini 
et al. 1991) have been used to compute the contributions of E/S0's and
Sp/Ir's in Fig. 2. Their sum fits very well the 6.7 $\mu m$
galaxy counts and also the K-band counts in Fig. 4.

On the other hand, the strong evolution required by the far-IR galaxy 
counts clearly concerns a different emission component, i.e. dust
illuminated by massive stars. The observed evolution is perhaps due to 
an increased rate of interactions (and merging?) in these gas-rich systems 
at redshifts $\simeq 1$.  Similar evolutionary effects are observed
at short optical wavelengths (the Faint Blue Galaxies, see e.g. Ellis 1997),
in the radio and perhaps in the X-ray band too.
We do not have yet a self-consistent physical description of this 
evolution of gas-rich late-type systems, we just modelled it with a 
kynematical increase of the luminosity as $L(z)=L(0)\ e^{2\tau(z)}$ in
an open universe ($q_0=0.15$) (see the short-dashed lines in Figs. 1, 3 
and 5).

Is there a conflict between this evidence of strong evolution in the FIR 
for gas-rich systems and their lack of evolution in the NIR?  This may not be 
the case if we consider that NIR observations sample the
integrated emission of aged stellar populations, and are then very weakly 
sensitive to evolution, while 
the FIR (and also the optical-UV and radio) sample
transient enhancements of massive-star formation due to interactions, more 
frequent in the past. These short-lived ($\Delta t\sim $ a few $10^7\ yr$)
starbursts may not add much to the mass of long-lived red stellar
populations observed in the NIR, where they could even be hardly observable.

We have also assumed in Figs. 1 to 5 that the early-type E/S0 systems 
evolve according to Franceschini et al. (1994), who suggested that a
bright phase of star formation and metal production 
at z$\sim$ 2 to 5 is obscured by dust
present in the ISM and quickly produced by the early stellar generations.
As it appears in the figures, given their high redshift and faintness, 
it is unlikely that these objects will be detected by ISO.

Finally, the contribution of AGNs in the FIR has been computed assuming 
the existence of a numerous population of self-absorbed objects advocated
to explain the X-ray background (see Granato et al. 1997 for more details).

\section
{CONSTRAINTS FROM THE IR BACKGROUND ON THE STAR FORMATION HISTORY} 

The difficulty to access faint distant sources in the IR, and 
the availability of two fairly clean spectral windows in the near-IR
and sub-mm, make the cosmic IR background an essential too to
investigate the distant universe.
Implementation of high-sensitivity detectors on space platforms (COBE)
have allowed to approach detection of the CIRB in both channels. 
Data and model predictions on the CIRB are summarized in Figure 7.
The lower dotted curve corresponds to the minimal expected contribution
from non-evolving galaxies, an estimate relying uniquely on the             
$\lambda$-dependent LLF (see Sect. 3.1). We may already see some confirmations
of our findings in the previous Sects.: a narrow margin available for 
evolution in the NIR, and wide room allowed by current limits in the
FIR/sub-mm.

\subsection{ The near-IR background} 

For $1\ \mu m<\lambda <8.5\ \mu m$ the galaxy contribution 
to the CIRB (thick line in Fig. 7) is estimated assuming the best-fit 
model of Figs. 2 and 4. 
The constraints set by very deep counts of galaxies in the K band and
recently by ISO at 6.7 $\mu m$ (Oliver et al. 1997) make this estimate
quite a robust one. 

 \begin{figure}[!ht]
  \begin{center}
    \leavevmode
    \centerline{\epsfig{file=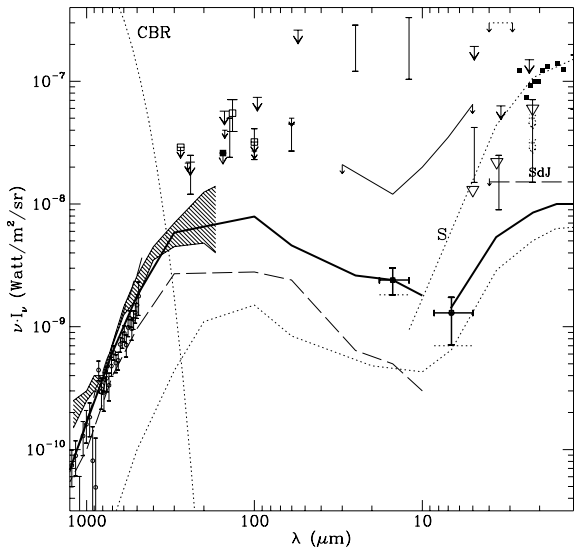,width=8.0cm}}
\vspace{-1.cm}
  \end{center}
  \caption{\em The extragalactic IR background, data versus model 
predictions. Data at the long-$\lambda$ part are from Puget et al. (1996) 
and Fixsen et al. (1997). Vertical bars are limits to the CIRB by Hauser 
(1996). Open triangles are upper limits by Kashlinsky
et al. (1996). The curve marked SdJ is from Stecker \& De Jager (1996) 
and the limits at 5 to 30 $\mu m$ are from Dwek \& Slavin (1994).
Curve S is the integrated starlight at the galactic pole.
The lower dotted line is the predicted flux for no evolution.
The dotted line is the partial contribution due to high-z dusty spheroids.
The two data-points at 6.7 and 15 $\mu m$ come from an evaluation of
the distant galaxy background based on the best-fit model discussed in
Sect. 3.3 and fitting available counts at these wavelengths. The 
corresponding lower dotted lines show the background intensity already
resolved by ISO-CAM into discrete sources.   See Sect. 4 for more details.
}
  \label{fig7}
\end{figure}

Observational limits, namely the Hauser (1996) and Stecker and de Jager 
(1996) estimates and the one by Kashlinsky et al. (1996) based on a
CIRB auto-correlation analysis, are already quite close to such a predicted
background. Altogether, the {\bf current information is
consistent with the NIR background mostly originating from ordinary
stellar populations in galaxies and very seriously limits any residual  
contributing signals from high-z processes} (e.g. pre-galactic stars,
emissions by primeval collapsed objects, decaying particles).

\subsection{ The FIR/SUB-MM background} 

Given a most favourable combination of the foregrounds, we could have had
perhaps here {\bf the first ever detection of the integrated emission 
of distant galaxies} in the form of an isotropic signal discovered by
Puget et al. (1996), whose spectrum is reported in Fig.7 at $200 <\lambda
<500\ \mu m$.
A discussion is now underway to ascertain that this signal is not due to,
e.g., a cold dust halo in our galaxy or in the solar neighbourhood.
Without entering the controversy, let us illustrate the possible impact 
of such a detection on the structure formation issue. 

Dust emission spectra for a wide range of galaxy
populations are all rather similar, due to the weak
dependence of dust equilibrium temperature on the radiation field 
intensity ($T_d \propto I_\nu^{1/5}$).
Then a knowledge of the CIRB spectral intensity at $\lambda \geq 100\ 
\mu m$ translates rather directly into a constraint on the effective
redshift $z_{eff}$ of sources making it.
Available data on the CIRB sub-mm spectrum constrain $z_{eff}$ to be
(Burigana et al. 1997):
\begin{equation} 
1+z_{eff} \simeq (3-5) \left({ T_d \over 50\ K} \right)
\end{equation} 
Similarly, the CIRB's observed intensity demands an energetics which is a 
function of the redshift $z_{eff}$ when the SF event occurred, of the
temperature $T_d$ of the re-processing dust, the fraction $F_{\rm FIRB}$
of the light 
re-processed into the IR, the efficiency $\epsilon$ of energy production 
by stellar processes. We can express this energy in terms of
the fraction $f$ of the total barions, as inferred from primordial 
nucleosynthesis, which have been processed by this high-z SF 
event to produce a solar metallicity:
\begin{equation} 
f \simeq 0.1 {0.05 \over \Omega_b} {Z_\odot\over Z} {1\over F_{\rm FIRB}} 
{0.007 \over \epsilon } \left({T_d\over 50\,K}\right)^{2.5} \left({4 
\over 1+z_{eff} }\right)^{1.5}, \label{eq:4}
\end{equation} 

(where it is assumed $\Delta Y = 2.5Z$ and $Z_\odot = 0.02$). This 
fraction is consistent with the barion mass and metallicity of local
early-type and cluster plasmas. 

As shown in Fig.7 (thick line), the model fitting FIR and sub-mm 
counts, assumed $T_d\simeq 50\ K$ for the intense star-burst phase at 
$z_{eff}\simeq 4$, 
predicts a far-IR/sub-mm background in close agreement with the Puget
et al. measurement. 
This is perhaps not as easy for typical models of hierarchical clustering 
based on the $q_0=0.5$ paradigm, which predict the main starburst 
phase occurring too late in cosmic time to reproduce the CIRB spectrum at 
long-wavelengths: the general tendency for these models may be to synthesize
excess flux at $\lambda\sim 100-200\ \mu m$, and much too little longwards. 

This stresses the importance of confirming, or disproving, the
FIR/sub-mm background, to gain a fairly direct test of -- otherwise
currently undetectable -- SF processes. 

 \begin{figure}[!ht]
  \begin{center}
    \leavevmode
    \centerline{\epsfig{file=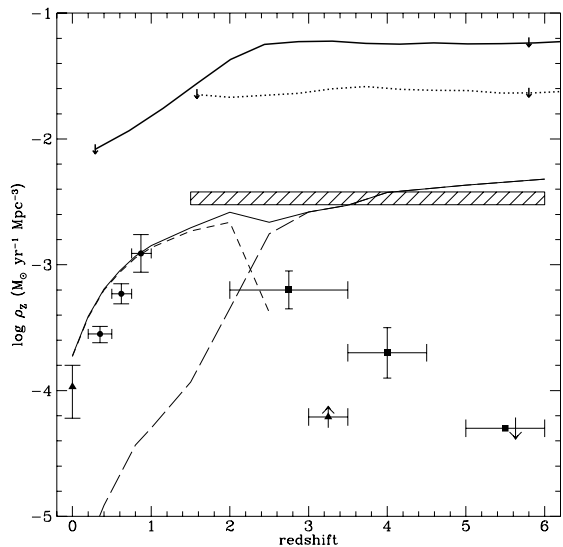,width=8.0cm}}
\vspace{-1.cm}
  \end{center}
  \caption{\em The history of the cosmic metal production rate.
Data-points are evaluations based on optical galaxy surveys: the CFRS 
survey at$z<1$ and on high-z galaxies selected with the "Lyman drop-out"
technique. The shaded region is an estimate by Mushotzky \& 
Loewenstein (1997) using as a constraint the metals observed in galaxy 
clusters.
Upper lines are limits based on a conservative analysis of COBE residuals
by Burigana et al. (1997). Long and short dashes correspond to the two
evolutionary branches of high-z dusty spheroids and low-z gas-rich
systems.
}
  \label{fig8}
\end{figure}

%
%
%
%
%
%
%

\section{\bf A TENTATIVE SCHEME FOR GALAXY EVOLUTION}

IR observations (and to some extent optical and radio observations too)
provide consistent indications in favour of a bimodal pattern for galaxy 
evolution:
\\
{\it (a)} the so-called early-type E/S0 systems, already quite old at z=1-2, 
evolve passively from that epoch to the present time; their formation  
should be confined to $z>2$;
\\ 
{\it (b)} late-type gas-rich disk-dominated systems make stars fairly 
actively at z=1-2 and below, i.e. over most of the Hubble time.

This scheme is illustrated in Figure 8, where
a lot of information on the history of star formation and metal production
is summarized. The two galaxy populations imply two different 
branches in this evolutionary history: the branch at $z<2$ 
corresponds to the phase of star and metal production in gas-rich systems, 
while the wide plateaux at z=2 to 6 marks the active, dust enshrouded phase 
of spheroid formation. 
Data points are estimates of the metal production rate based
on optical observations: it is clear that, according to these data,
the high-redshift branch should be mostly unobservable in the optical,
consistent with the view that most of the light was dust-reprocessed 
into the IR/sub-mm.

In addition to the above discussed indications for a high-redshift phase 
of excess SF based on the old field galaxies observed at high-z
(Sects. 3.2 and 3.3)
and on the requirements set by the Cosmic IR/sub-mm Background (Sect. 4.2),
independent evidences come from:

i) the early-type galaxy populations in galaxy clusters, passively 
evolving to $z=1.2$ (e.g. Dickinson 1997), the large amounts of metals 
and the lack of evolution in the temperature and density of the 
Intra-Cluster Medium up to $z=0.7$ (Mushotzky \& Loewenstein, 1997); 
{\bf note in particular the impressive agreement in Fig. 
8 between the estimated rate of production of metals in the ICM and our
curve needed to reproduce the sub-mm background};

ii) the copious amounts of dust observed in high- and very high-z QSO's
(see Andreani et al. [these Proceedings]);

iii) the amounts of metals in the environments of high-z QSO's (as shown by
the associated absorbers, see Franceschini \& Gratton, 1997).

A major episode of star formation at $z>2$ has obvious significant 
implications for structure formation scenarios. Combined with 
clues of the existence of massive clusters already in place at 
$z>>1$ from SZ observations (Richards et al. 1997),
with direct evaluations of $\Omega$ based on measurements of
the global $M/L$ of galaxy clusters and its redshift evolution
(Carlberg et al. 1997), and with the observed 
barion fractions in clusters, then all this may require modification
of the gravitational instability picture based on the conventional  
$\Omega$=1-CDM paradigm.

\section{\bf OPEN PROBLEMS, AND WHO COULD CONTRIBUTE TO SOLVE THEM}

We believe that disprove or substantiation of the tentative scheme of
galaxy formation and evolution outlined in Sect. 5, versus other
alternative interpretations, will be one of the crucial motivations of 
observational cosmology in the next several years. The present Section 
is devoted to a short discussion of three related open questions, with 
the aim of clarifying the opportunities and limitations of planned facilities.

\subsection{Properties of the evolving population of 
gas-rich late-types and irregulars at $0<z<2$}

Essentially in any accessible wavebands (apart the near-IR), 
populations of active galaxies have been discovered with
perhaps similar properties to those of gas-rich systems found in the
IR: 

{\it a)} in the optical-UV the Faint Blu Objects dominating the 
short-wavelength optical counts at faint magnitudes (see e.g. Ellis 
1997);

{\it b)} in the radio centimetric, the milli-Jy to micro-Jy radio galaxy 
population responsible for the upturn of the radio counts;

{\it c)} the narrow emission-line galaxies associated with faint X-ray ROSAT 
sources at $S \leq 10^{-14}\ erg/cm^2/s$ (McHardy et al., 1997).

 \begin{figure}[!ht]
  \begin{center}
    \leavevmode
    \centerline{\epsfig{file=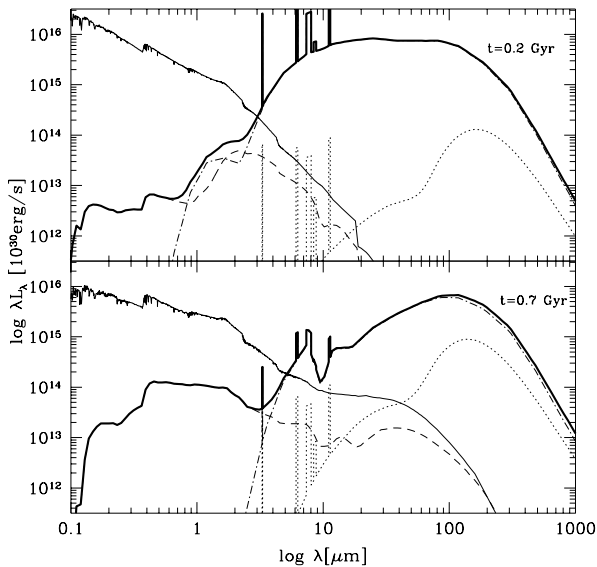,width=8.0cm}}
\vspace{-1.cm}
  \end{center}
  \caption{\em Predicted rest-frame spectra at two ages (0.2 and 0.7 Gyr) 
of a star-forming galaxy with $M=5\ 
10^{11}\ M_\odot$, a decay time for the SF of $\tau=0.5$ Gyr, forming stars 
at rates of 700 and 400 $ M_\odot/yr$ at the two respective ages, 
and producing a galactic wind at an age of 0.75 Gyr. 
The thin continuous line is the optical unabsorbed spectrum, the dashed
line the absorbed one, the dot-dashed the emission by molecular
clouds and the dotted line is the cold "cirrus" emission. The fractions of 
residual gas are 0.7 and 0.4, while the dust/gas is 0.002 and 0.009 
(proportional to the metallicity), respectively. Star-forming regions 
are modelled as molecular clouds, assumed to include 30\% of all 
residual gas at any moment
(see and Granato et al., these Proceedings, for more details). }
  \label{fig9}
\end{figure}

The relationships among these various populations of active and 
starbursting galaxies are at the moment quite unclear.
What is indicated by existing data is that there is no template spectrum
for the whole population. The Spectral Energy Distribution (SED's) may 
dramatically depend on the galaxy's morphological type, selection wavelength,
age of the object and cosmic epoch.
Points to be understood would be, in particular:

a) does the IR-selected evolving population of starburts coincide with
the moderate-to-low mass systems dominating the blue optical counts?, or

b) is it a somewhat unrelated population of more massive objects, 
strongly interacting and probably merging, like the hyperluminus IR galaxies?

c) what is the origin of the strong observed evolution with redshift up
to at least $z=1$? 

d) what is the amount of barions locked into stars in this relatively 
late phase? Note that the mass in stars might not be strictly
related to the amount of metals appearing in Fig. 8, if e.g. the IMF 
varies.

FIRST will characterize the dust emission spectrum in a 
critical range ($\sim 80$ to 600 $\mu m$) where peak emission is 
expected to be observable. Figure 9 reports the synthetic spectrum 
of a massive starburst galaxy observed after
0.2 and 0.7 Gyr from the onset of the SF. FIRST's
spectral coverage will then sample the redshifted photons bringing the 
bulk of the energy during a sizeable fraction of galaxy's lifetime.
Given the moderate redshifts this phenomenon is occurring at, 
important contributions will also come from post-ISO
moderate-size IR missions, like SIRTF, sampling the critical window,
from a few $\mu m$ to a few tens of $\mu m$, where hot dust
in the star-forming regions is emitting. FIRST and planned IR space 
facilities are very nicely complementary on this respect.

 \begin{figure}[!ht]
  \begin{center}
    \leavevmode
    \centerline{\epsfig{file=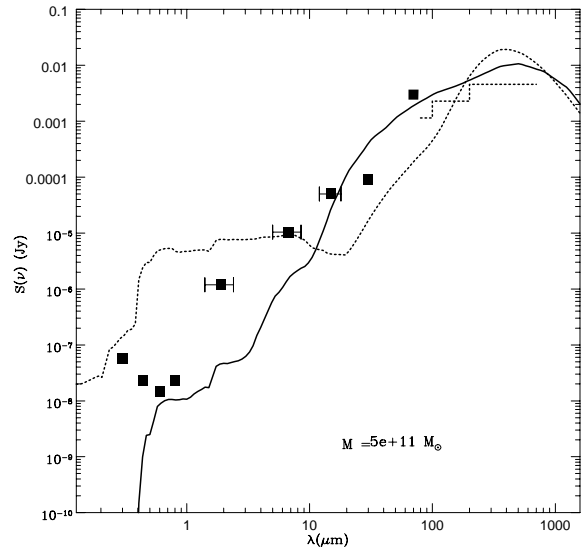,width=8.0cm}}
\vspace{-1.cm}
  \end{center}
  \caption{\em Predicted spectrum in the observable plot (flux versus 
observed wavelength) of a massive primeval star-forming galaxy 
with z=3.5 (see text). The rest-frame 
spectra are those reported in Fig. 9.   These predictions are compared 
with sensitivities for long integrations (3 hours) of, from left to right, 
HST Hubble Deep Field in U, B, V, I,
K=22, ISO-CAM LW2 and LW3 and SIRTF at 30 and 70 $\mu m$. 
The dotted step-like line is the predicted sensitivity of FIRST. Note that
the SIRTF long-wavelength sensitivity is limited by confusion noise.
}
  \label{fig10}
\end{figure}

\subsection{Formation and early-evolution of AGNs and QSOs}

We have seen that the evolution rates of IR-selected starburst populations
are comparable to those of optical and X-ray quasars 
($L[z] \propto [1+z]^{2-3}$). Basic questions are:

i) How the starburst is related to the AGN phenomenon? Are
evolving luminous and hyper-luminous IR galaxies primarily energized 
by dusty AGNs?

ii) How super-massive Black Holes (of $M \sim 10^8-10^9\ M_\odot$) are
formed in quasar nuclei at high-z ?
It is common wisdom that this event occurr during early formation of
spheroids, probably in a dusty medium, while the classical
optical quasar phase would be a late one in the process of QSO 
formation.

iii) A question more related to local AGN phenomenon, and in particular
to the unified model of AGN activity:
how many type-II AGN do exist, and how do they evolve?

Answers are likely to come from IR studies. In particular,
if (iii) will be addressed also by other missions, points (i) and (ii) will 
probably require FIRST, because of the redshift effect.

\subsection{Evolutionary history of galaxies at $z>2$}

Finally, we consider the class of early-type E/S0 galaxies mentioned 
in the evolutionary scheme of Sect. 5.
The passively evolving, old stellar populations observed at 1 to 8.5
$\mu m$ by ISO and by K-band ground-based surveys, in both field and 
cluster galaxies, require that the bulk of the Star Formation making 
them should have occurred at $z>2$. This star-formation history has 
been modelled in Fig. 8 as a wide plateaux, declining at $z<2$.
 
Also, if we take the metal abundance measured in X-rays
in the Intra-Cluster Medium (assuming it sinthetized by cluster 
galaxies) as representative of the whole E/S0 population, then
the implication might be that the high-redshift SF phase should have
involved a massive-star and metal production in excess of the
lower-redshift SF branch in Fig. 8.

If all this occurred in a dusty medium (as suggested by lack of detection
in the optical, and, on the positive side, by the observed long-wavelength 
excess in the CIRB), IR surveys are needed to detect it.
In this case the redshift should be quite high, and the peak emission 
would be expected to fall at $\lambda>100\ \mu m$, hence best 
(or even only) observable by FIRST among the planned IR space missions. 

Figure 10 matches the sensitivity for long integrations of various 
facilities with model spectra of a major starburst at $z=3.5$, 
corresponding to the early phase of top Fig. 9. The two curves 
differ only for the fraction of gas in the molecular phase, which is
30\% for the continuous line and 0\% for the dotted line.
While the optical to far-IR 
spectrum is much dependent on the age of the object, plus a number of 
details such as the average dust optical depth, the physical parameters 
of the star-forming regions, how distant is the dust from the 
illuminating sources, and so on, the sub-millimeter spectrum is 
much less affected by these uncertain quantities and implies a robust 
test for the existence of an early dust-enshrouded phase.

Altogether, FIRST promises some fundamental progresses in our knowledge of 
the high-z universe.
From $\lambda=80$ to several hundreds $\mu m$ it appears as a unique
instrument for astronomical exploration, 
even if compared with the most ambitious projects now under scrutiny.
In particular, performances of FIRST and of the New-Generation Space 
Telescope look very much complementary and synergic. 
FIRST's wavelength range is perhaps where
peak emission from forming structures is to be expected. 
       
%

\end{document}